\documentclass[conference]{IEEEtran}
\IEEEoverridecommandlockouts
\usepackage{cite}
\usepackage{amsmath,amssymb,amsfonts}
\usepackage{algorithmic}
\usepackage{graphicx}
\usepackage{subfigure}
\usepackage{textcomp}
\usepackage{xcolor}
\def\BibTeX{{\rm B\kern-.05em{\sc i\kern-.025em b}\kern-.08em
    T\kern-.1667em\lower.7ex\hbox{E}\kern-.125emX}}

\usepackage{makecell}

\usepackage{acro}
\DeclareAcronym{3GPP}{
  short=3GPP,
  long=3rd generation partnership project
}
\DeclareAcronym{ADC}{
  short=ADC,
  long=analog-to-digital converter
}
\DeclareAcronym{AMP}{
  short=AMP,
  long=approximate message passing
}
\DeclareAcronym{ANN}{
  short=ANN,
  long=artificial neural network
}
\DeclareAcronym{AoA}{
  short=AoA,
  long=angle-of-arrival
}
\DeclareAcronym{AoD}{
  short=AoD,
  long=angle-of-departure
}
\DeclareAcronym{APS}{
  short=APS,
  long=azimuth power spectrum
}
\DeclareAcronym{AR}{
  short=AR,
  long=augmented reality
}
\DeclareAcronym{AV}{
  short=AV,
  long=autonomous vehicle
}
\DeclareAcronym{BM}{
  short=BM,
  long=beam management
}
\DeclareAcronym{BS}{
  short=BS,
  long=base station
}
\DeclareAcronym{BSM}{
  short=BSM,
  long=basic safety message
}
\DeclareAcronym{BW}{
  short=BW,
  long=bandwidth
}
\DeclareAcronym{CDF}{
  short=CDF,
  long=cumulative distribution function
}
\DeclareAcronym{CP}{
  short=CP,
  long=cyclic-prefix
}
\DeclareAcronym{CSI-RS}{
  short=CSI-RS,
  long=channel state information reference signal
}
\DeclareAcronym{DFT}{
  short=DFT,
  long=discrete Fourier transform
}
\DeclareAcronym{DL}{
  short=DL,
  long=downlink
}
\DeclareAcronym{EKF}{
  short=EKF,
  long=extended Kalman filter
}
\DeclareAcronym{DSRC}{
  short=DSRC,
  long=dedicated short-range communication
}
\DeclareAcronym{FDD}{
  short=FDD,
  long=frequency division duplex
}
\DeclareAcronym{FMCW}{
  short=FMCW,
  long=frequency modulated continuous wave
}
\DeclareAcronym{FoV}{
  short=FoV,
  long=field-of-view
}
\DeclareAcronym{GNSS}{
  short=GNSS,
  long=global navigation satellite system
}
\DeclareAcronym{IMU}{
  short=IMU,
  long=inertial measurement unit
}
\DeclareAcronym{IC}{
  short=IC,
  long=integraed circuit
}

\DeclareAcronym{lidar}{
  short=lidar,
  long=light detection and ranging
}
\DeclareAcronym{LOS}{
  short=LOS,
  long=line-of-sight
}
\DeclareAcronym{LPF}{
  short=LPF,
  long=low pass filter
}
\DeclareAcronym{LTE}{
  short=LTE,
  long=long term evolution
}
\DeclareAcronym{MIMO}{
  short=MIMO,
  long=multiple-input multiple-output
}
\DeclareAcronym{ML}{
  short=ML,
  long=machine learning
}
\DeclareAcronym{mmWave}{
  short=mmWave,
  long=millimeter wave
}
\DeclareAcronym{MRR}{
  short=MRR,
  long=medium range radar
}
\DeclareAcronym{NLOS}{
  short=NLOS,
  long=non-line-of-sight
}
\DeclareAcronym{NB}{
  short=NB,
  long=narrow beam
}
\DeclareAcronym{NR}{
  short=NR,
  long=new radio
}
\DeclareAcronym{OFDM}{
  short=OFDM,
  long=orthogonal frequency-division multiplexing
}
\DeclareAcronym{ppm}{
  short=ppm,
  long=parts-per-million
}
\DeclareAcronym{PF}{
  short=PF,
  long=particle filter
}
\DeclareAcronym{PUCCH}{
    short=PUCCH,
    long=physical uplink control channel
}
\DeclareAcronym{PUSCH}{
    short=PUSCH,
    long=physical uplink shared channel
}
\DeclareAcronym{RMS}{
  short=RMS,
  long=root-mean-square
}
\DeclareAcronym{RPE}{
  short=RPE,
  long=relative precoding efficiency
}
\DeclareAcronym{RS}{
  short=RS,
  long=reference signal
}
\DeclareAcronym{RSRP}{
  short=RSRP,
  long=reference signal received power
}
\DeclareAcronym{RSU}{
  short=RSU,
  long=roadside unit
}
\DeclareAcronym{SCS}{
  short=SCS,
  long=subcarrier spacing
}
\DeclareAcronym{SNR}{
  short=SNR,
  long=signal-to-noise ratio
}
\DeclareAcronym{SSB}{
  short=SSB,
  long=synchronization signal block
}
\DeclareAcronym{THz}{
  short=THz,
  long=terahertz
}
\DeclareAcronym{TTD}{
  short=TTD,
  long=true-time-delay
}

\DeclareAcronym{UAV}{
  short=UAV,
  long=unmanned aerial vehicle
}
\DeclareAcronym{UE}{
  short=UE,
  long=user equipment
}
\DeclareAcronym{UKF}{
  short=UKF,
  long=unscented Kalman filter
}
\DeclareAcronym{UL}{
  short=UL,
  long=uplink
}
\DeclareAcronym{ULA}{
  short=ULA,
  long=uniform linear array
}
\DeclareAcronym{V2I}{
  short=V2I,
  long=vehicle-to-infrastructure
}
\DeclareAcronym{V2V}{
  short=V2V,
  long=vehicle-to-vehicle
}
\DeclareAcronym{V2X}{
  short=V2X,
  long=vehicle-to-everything
}
\DeclareAcronym{VR}{
  short=VR,
  long=virtual reality
}
\DeclareAcronym{VRU}{
  short=VRU,
  long=vulnerable road user
}
\DeclareAcronym{WB}{
  short=WB,
  long=wide beam
}
\DeclareAcronym{RNN}{
	short=RNN,
	long=recurrent neural network
}
\DeclareAcronym{LSTM}{
	short=LSTM,
	long=long short-term memory
}

\DeclareAcronym{FC}{
	short=FC,
	long=fully connected
}

\DeclareAcronym{PC}{
	short=FC,
	long=partially connected
}

\DeclareAcronym{JPTA}{
    short=JPTA,
    long = joint phase-time array
}
\DeclareAcronym{PAA}{
    short=PAA,
    long = phased antenna arrays
}

\usepackage{amsmath,amssymb}

\newcommand{\bb}{{\mathbf{b}}}

\newcommand{\bd}{{\mathbf{d}}}

\newcommand{\bp}{{\mathbf{p}}}

\newcommand{\bT}{{\mathbf{T}}}

\makeatletter
\def\munderbar#1{\underline{\sbox\tw@{$#1$}\dp\tw@\z@\box\tw@}}
\makeatother

\newcommand{\figref}[1]{Fig.~\ref{#1}}

\DeclareMathOperator*{\argmin}{arg\,min}

\begin{document}

\title{Beamforming with Joint Phase and Time Array: System Design, Prototyping and Performance}

\author{\IEEEauthorblockN{Jianhua Mo, Ahmad AlAmmouri, Shenggang Dong, Younghan Nam, \\ Won-Suk Choi, Gary Xu, and Jianzhong (Charlie) Zhang}
\IEEEauthorblockA{\textit{Standards and Mobility Innovation Laboratory, Samsung Research America}, Plano, TX 75024, USA}
\{jianhua.m, ahmad1.a, s.dong, younghan.n, wswill.choi, gary.xu, jianzhong.z\}@samsung.com}

\maketitle

\begin{abstract}
Joint phase-time arrays (JPTA) is a new mmWave radio frequency front-end architecture constructed with appending time-delay elements to phase shifters for analog beamforming. JPTA allows the mmWave base station (BS) to form multiple frequency-dependent beams with a single RF chain, exploiting the extra degrees of freedom the time-delay elements offer. Without requiring extra power-hungry RF chains, a BS with JPTA can schedule multiple users in different directions in a frequency-division multiplexing (FDM) manner. A BS with JPTA achieves various advantages over the traditional analog beamforming system.
Simulation results show that JPTA can bring significant system-level benefits, e.g., extending uplink throughput coverage by 100\%. To realize these system benefits of JPTA, high-resolution delay elements with a wide delay dynamic range are essential. With newly developed delay elements, we demonstrate that a single TRX RF chain can serve four users in four different directions in the mmWave band.
\end{abstract}


\maketitle

\section{Introduction}
\label{sec:introduction}

Millimeter-wave (mmWave) technology represents a pivotal advancement in 5G cellular systems, harnessing the vast potential of the mmWave spectrum to deliver unprecedented data rates. This technology capitalizes on the abundant yet underutilized bandwidth in the mmWave band, promising significantly enhanced network performance and user experiences. However, despite its immense potential, mmWave technology comes with challenges primarily stemming from the high pathloss characteristics inherent to mmWave frequencies. The propagation characteristics of mmWave signals are characterized by high attenuation and susceptibility to environmental obstacles, such as buildings and foliage, which significantly limit the coverage area of cellular networks operating in this frequency range.

To overcome the limitations imposed by high pathloss, extensive research efforts have been dedicated to developing innovative solutions for enhancing coverage and improving the overall efficiency of mmWave-based cellular systems. One of the key strategies adopted in 5G standards is utilizing large-antenna narrow beamforming techniques \cite{Six_Heng21}. By employing large antenna arrays and focusing transmission beams into narrow beamwidth, beamforming enables the concentration of signal power towards desired users or areas, thereby mitigating the adverse effects of pathloss and extending the reach of mmWave networks.

Despite the effectiveness of beamforming in enhancing coverage, the implementation of large antenna arrays presents challenges, particularly in terms of power consumption and hardware costs. Fully digital beamforming can incur significant power consumption and hardware expenses while offering precise control over beamforming parameters. In response to these challenges, analog beamforming has emerged as a cost-effective alternative, offering a simplified architecture that reduces the complexity and hardware requirements associated with fully digital implementations.

Analog beamforming systems typically consist of a single digital transceiver unit (TRX or RF chain) connected to an antenna panel housing multiple antennas. However, the adoption of analog beamforming introduces what is known as the “analog beamforming constraint,” limiting the system to utilizing only a single analog beam across the entire system bandwidth within a given time duration. This constraint can lead to inefficient utilization of available bandwidth, increased scheduling latency, and decreased user throughput, particularly in scenarios with varying traffic demands and channel conditions.

To address the limitations of analog beamforming and enhance the flexibility of beamforming operations, hybrid analog-digital beamforming (HBF) techniques have been proposed \cite{Ayach_TWC14, Alkhateeb_MCOM14, Heath2016, Molisch_HP_mag}. These techniques involve the incorporation of additional TRXs and phase shifter arrays, allowing for more sophisticated beamforming capabilities while minimizing the associated costs and power consumption. However, integrating extra hardware components introduces challenges, including increased complexity and signal processing requirements, which can impact system performance and scalability.

In light of these challenges, there is a growing interest in exploring alternative beamforming architectures that balance cost-effectiveness and performance. One such architecture is the joint phase-time array (JPTA), which integrates true-time-delay (TTD) units alongside traditional phase shifters and switches used in analog and hybrid beamforming \cite{Joint_Vishnu22,alammouri_extending_2022, yildiz20243dbeamformingjointphasetime}. By leveraging TTD units to introduce frequency-dependent phase shifts, JPTA offers enhanced flexibility in beamforming operations, enabling the creation of optimized beam patterns tailored to specific environmental conditions and user requirements. Moreover, compared to conventional approaches involving multiple TRXs, JPTA presents a more cost-effective solution for enabling multi-beam operations, making it an attractive option for future mmWave-based cellular systems. JPTA is also called TTD beamforming, delay-phased array (DPA), delay phase precoding (DPP), and TTD-based hybrid precoding in the literature \cite{Delay_Dai2021,SS_Zhai2021,A_Lin2021, jain_mmflexible_2023, Fast_Boljanovic2021}.

The rest of the paper is organized as follows: Section \ref{sec:JPTA} discusses the JPTA architecture and provides two critical types of JPTA beams. Section \ref{sec:Prototyping} presents the prototyping of JPTA beamforming, which serves four users simultaneously. The simulation setup and the simulation results are discussed in Section \ref{sec:Simulation} before concluding in Section \ref{sec:Conclusion}.

\section{JPTA Beamforming}\label{sec:JPTA}
In this section, we introduce the JPTA architecture, the beam design of JPTA, and its comparison with HBF. 

\begin{figure}[t]
	\centerline{\includegraphics[width=  0.9\linewidth]{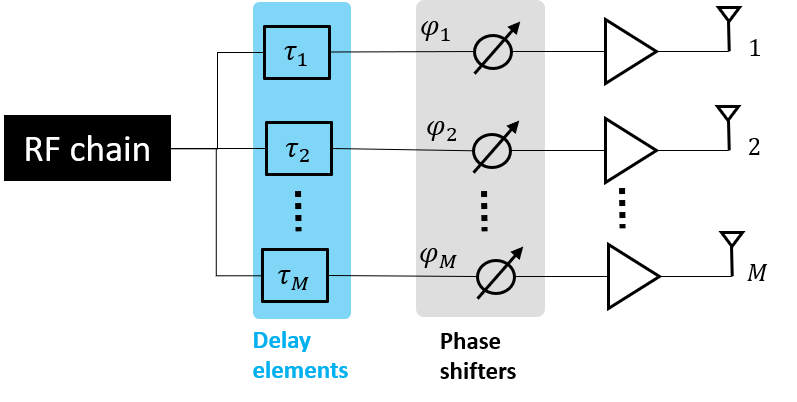}}
	\caption{The considered JPTA architecture, where each antenna element is connected through a delay unit and a PS.}
	\label{fig:JPTAArch}
\end{figure}

\subsection{JPTA Architecture} \label{subsec:JPTA_Arch}
The JPTA architecture is shown in \figref{fig:JPTAArch}. Each antenna is connected through an adjustable time delay (TD) element and an adjustable phase shifter (PS). Then all the antenna branches are connected through a single TRX (or called RF chain). Compared to the traditional phased antenna array, the key difference is the extra delay elements. Mathematically, the $M \times 1$ downlink transmitted signal on sub-carrier $k$ is,

\begin{align}
\mathbf{x}_k &= \underbrace{\frac{1}{\sqrt{M}} 
\begin{bmatrix}
    e^{{\mathrm{j}} \phi_{1}} & 0 & \hdots & 0 \\
    0 & e^{{\mathrm{j}} \phi_{2}} & \hdots & 0 \\
    \vdots & \vdots & \ddots & \vdots \\
    0 & 0 & \hdots & e^{{\mathrm{j}} \phi_{M}} 
  \end{bmatrix} }_{\mathbf{T}}
\underbrace{\left[\begin{array}{c} e^{{\mathrm{j}} 2 \pi f_k \tau_{1}} \\ e^{{\mathrm{j}} 2 \pi f_k \tau_{2}} \\ \vdots \\ e^{{\mathrm{j}} 2 \pi f_k \tau_{M}}\end{array} \right]}_{\mathbf{d}_k} s_k \\
&= \underbrace{ \frac{1}{\sqrt{M}} \begin{bmatrix}
    e^{{\mathrm{j}} \phi_{1} + 2\pi f_k \tau_1}  \\
    e^{{\mathrm{j}} \phi_{2} +2 \pi f_k \tau_2}  \\
    \vdots \\
    e^{{\mathrm{j}} \phi_{M} + 2\pi f_k \tau_M} 
  \end{bmatrix} }_{\mathbf{p}_k} s_k, \label{eqn_tx_signal}
\end{align}

\noindent where $\phi_m$ is the phase of the $m$-th phase shifter, $\tau_m$ is the delay of the $m$-th delay element, and $s_k$ is the modulation symbol on the $k$-th sub-carrier $f_k$. $\bT$ stands for the phase shifts applied to the $M$ antennas, and is independent of the frequency. $\bd_k$ represents the phase drift caused the delay elements at the frequency $f_k$. The phase drift, i.e., $2 \pi f_k \tau_m$, is frequency-dependent. The aggregated phase change in JPTA is denoted as $\bp_k$, which is also frequency-dependent.

\subsection{JPTA Beam Generation} \label{subsec:jpta_beam_gen}

\begin{figure}[t]
        \centering
        \subfigure[Type-1 JPTA Beam: Discrete-angle beam]{
            \includegraphics[width= 0.9\linewidth]{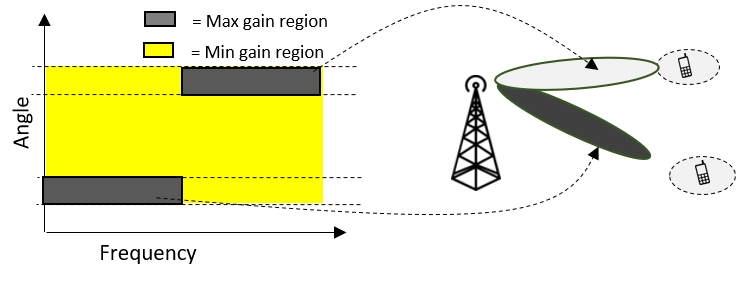}
            \label{fig:Type1_JPTA}}
        \subfigure[Type-2 JPTA Beam: Rainbow beam or continuous-angle beam]{
            \includegraphics[width= 0.9\linewidth]{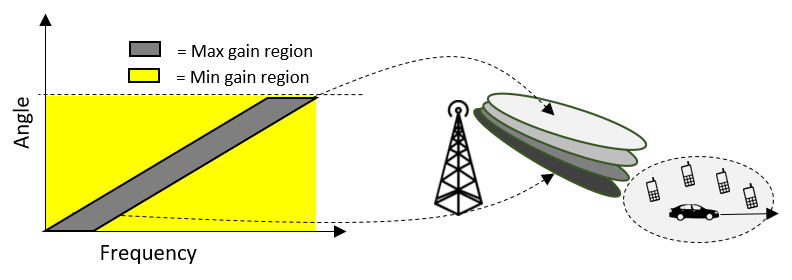}
            \label{fig:Type2_JPTA}}
         \caption{A simple illustration of Type-1 and Type-2 JPTA Beams.}
        \label{fig:JPTA_illustration}
\end{figure}

Two types of JPTA beam patterns shown in \figref{fig:JPTA_illustration} are considered in this paper. 

\subsubsection{Type-1 JPTA Beam}
Type-1 beam pattern refers to steering in multiple distinct, discrete directions. In an example shown in \figref{fig:Type1_JPTA}, the whole frequency band is split into two subbands steered to two users separately.
Denote $\bb_k$ as the desired beamforming weight vector at k-th sub-carrier. For a uniform linear array with antenna spacing $d$, the steering vector pointing to the direction $\theta_k$ at the frequency $f_k$ is,
\begin{align}
\bb_k = \frac{1}{\sqrt{M}} \left[ 1, \ e^{\mathrm{j} \frac{2 \pi d f_k \cos \theta_k}{c}}, \cdots,  \ e^{\mathrm{j} \frac{2 \pi (M-1) d f_k \cos \theta_k}{c}}  \right]^T
\end{align}
where $c$ is the speed of light.

We formulate an optimization problem to minimize the total difference between $\bp_k$ and $\bb_k$ over all the sub-carriers:
\begin{align}
\mathbf{\tau}, \mathbf{\phi} = \argmin_{\left\{\tau_m, \phi_m \right\}} \sum_{k} \lVert \mathbf{p}_k- \mathbf{b}_k \rVert^2.
\end{align}

The optimization problem can be solved by an iterative algorithm proposed in \cite{Joint_Vishnu22} or a heuristic single-shot solution in \cite{jain_mmflexible_2023}. Type-1 beam can be useful for scheduling a set of desired beams along with desired subband sizes in a given time slot. Examples of Type-1 beam patterns will be shown in Section \ref{sec:SimResDis}.

\subsubsection{Type-2 JPTA Beam}
Type-2 beam is called rainbow \cite{wadaskar_3d_2021, Rainbow_Li22} or prism beam \cite{THzPrism_Zhai2020}, which covers a specific beam spread contiguously across the bandwidth. If a JPTA beam is required to cover an angular range $(\theta_c- \frac{\Delta \theta}{2}, \theta_c + \frac{\Delta \theta}{2})
$, a closed-form solution of JPTA beam design is \cite{Rainbow_Li22,Joint_Vishnu22 } 
\begin{align}
\tau_m &= \frac{m-1}{W} \sin \frac{\Delta \theta}{2}, \\
\phi_m &=\frac{2 \pi (m-1) d \cos \theta_c}{\lambda}.
\end{align}

The potential use of Type-2 beams is to speed up the beam alignment by sweeping all the directions within one shot \cite{Rainbow_Li22}. It can also support high mobility users by re-transmitting the time-sensitive messages in the adjacent angles of the current serving beam. An example is shown in \figref{fig:Type2_JPTA}.

The Type-2 beam can be treated as a special case of type-1 when the Type-1 beam is designed to cover multiple discrete angles with slight separations. Therefore, we will focus on Type-1 beam in this paper.

\subsection{Comparison with Hybrid Precoding}
In traditional hybrid beamforming, analog components exhibit a frequency-flat response, while frequency-dependent beamforming is implemented digitally through multiple RF chains. A pertinent inquiry arises regarding the quantity of TRXs needed to replicate the beam patterns achievable by conventional hybrid beamforming.

Our research indicates that emulating JPTA's beam pattern requires many RF chains [4]. Specifically, for Type-1 beam patterns, the minimum number of TRXs should at least match the number of subbands. For Type-2 patterns, this number grows linearly with the increase in antenna count and angular range. Notably, the power consumption and cost of TTD elements are significantly lower than those of TRXs, making JPTA a more energy-efficient and cost-effective alternative.

\section{Prototyping} \label{sec:Prototyping}

\begin{figure*}[h!]
    \centering
    \includegraphics[width= 0.8\linewidth]{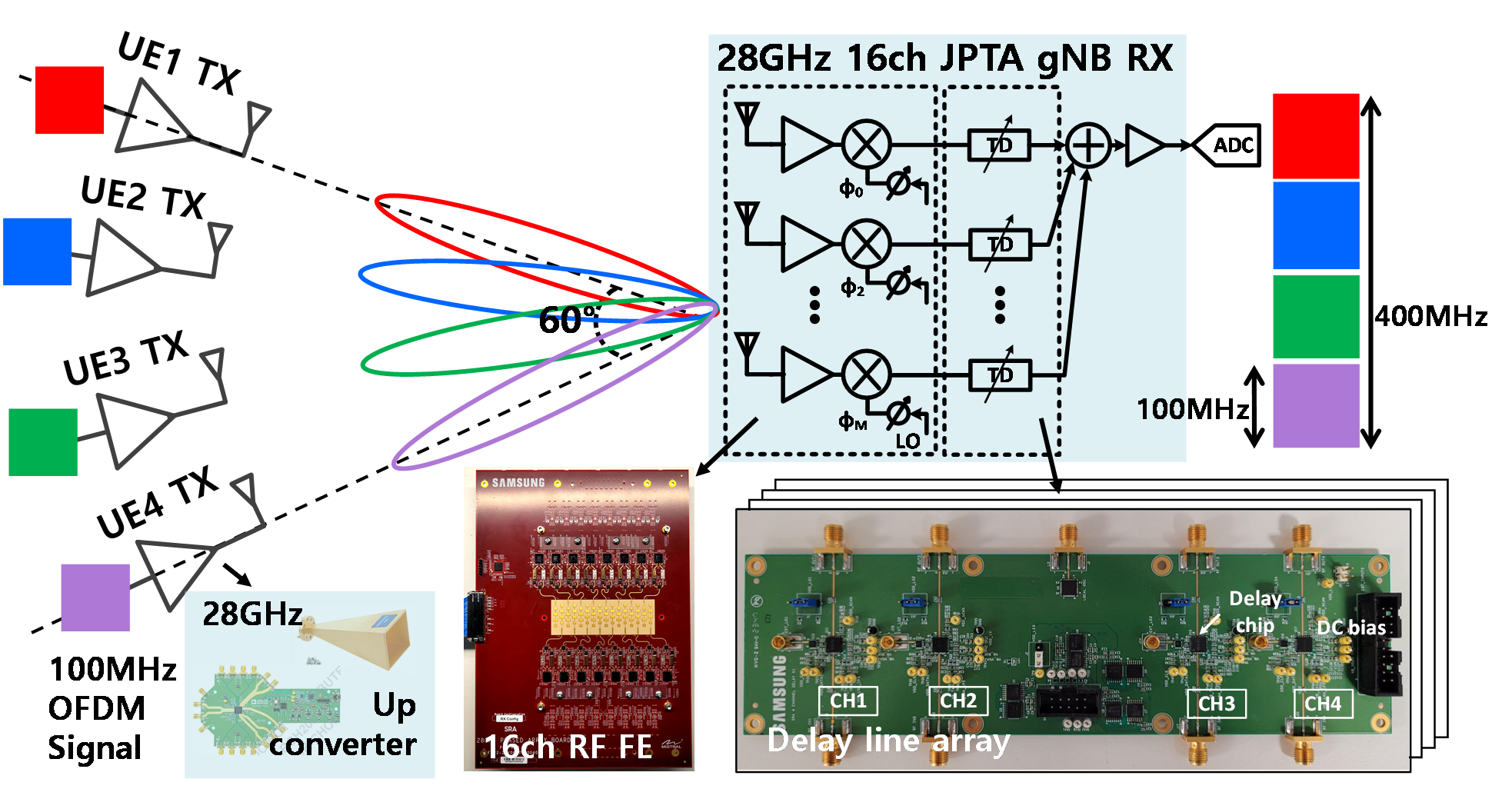}
    \label{fig:demoSetup}
    \caption{JPTA demonstration setup}
\end{figure*}

\begin{figure*}[h!]
    \centering
    \includegraphics[width= 0.8\linewidth]{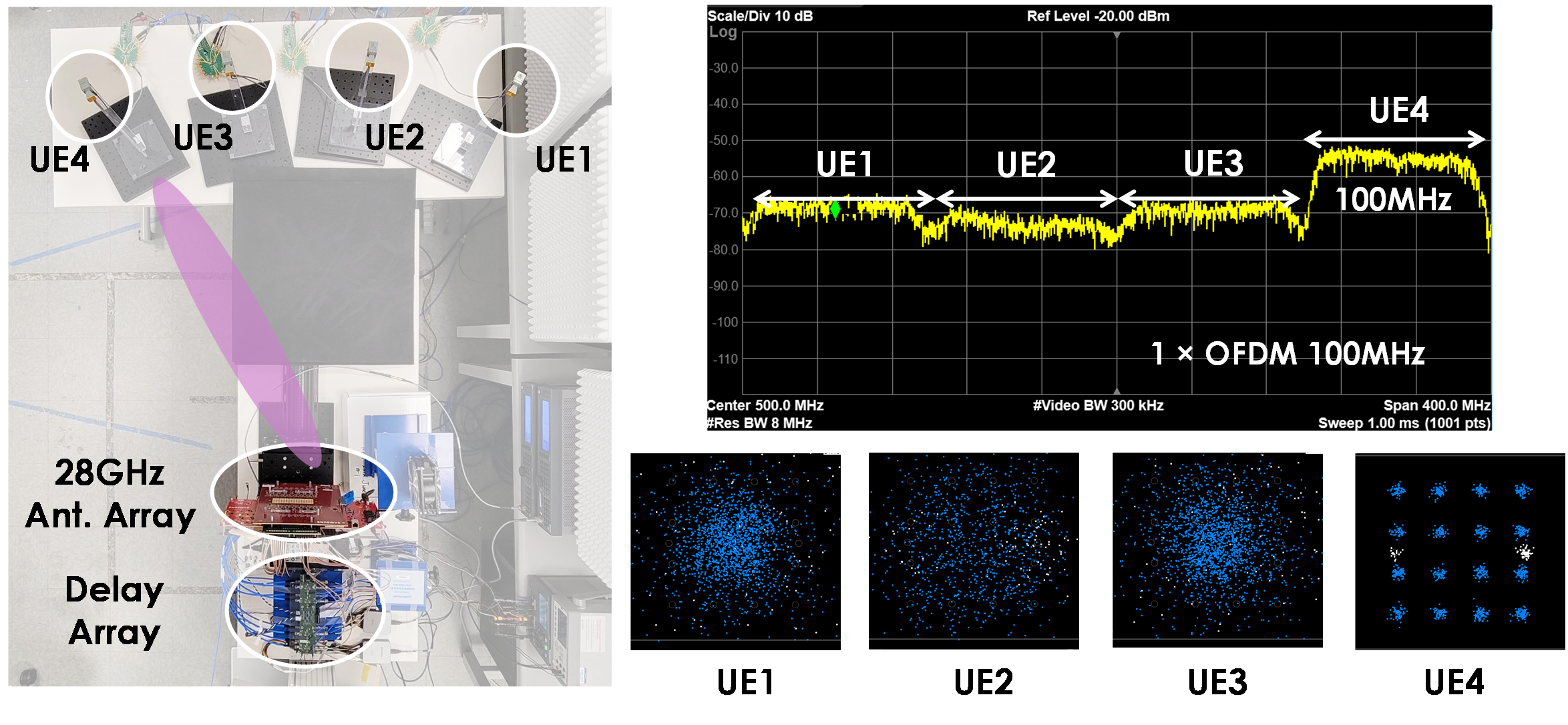}
    \label{fig:demoJPTAoff}
    \caption{JPTA off. BS serves 4 users sequentially. This figure shows when a beam steered to UE4.}
\end{figure*}

\begin{figure*}[h!]
    \centering
    \includegraphics[width= 0.8\linewidth]{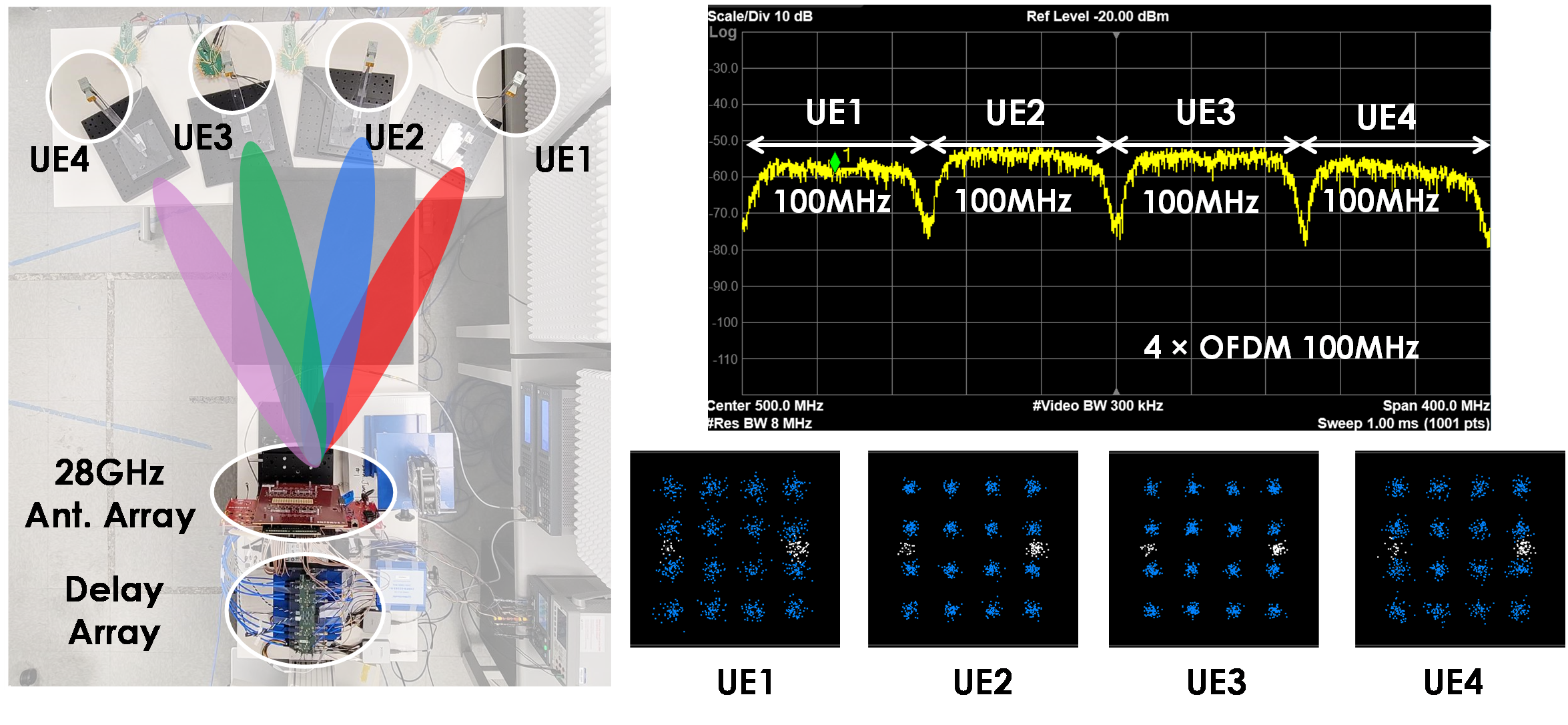}
    \label{fig:demoJPTAon}
    \caption{JPTA on. Four UEs are supported simultaneously.}
\end{figure*}

To verify the JPTA concept, a demonstration prototype for the uplink, which consists of one 28 GHz gNB and four UEs, is built, as shown in \figref{fig:demoSetup}. The goal is to demonstrate the feasibility of the JPTA frequency-dependent multi-beams feature using type-1 codebooks. 

An off-the-shelf upconverter (ADMV1018) followed by a 28 GHz horn antenna is adopted as the UE in the transmitting (TX) mode. In this setup, 4 UEs are located evenly across the 60° angular spread from the gNB receiver point of view. Assuming the system bandwidth is 400 MHz, each UE is set to send the 100 MHz OFDM signals to occupy the entire system bandwidth to demonstrate the extreme case. On the gNB receiver (RX) side, 16 RF channels are integrated onboard in the horizontal direction. Each RF channel is connected to one $1 \times 4$ series-fed patch antenna and down-converted to the 0.5 GHz intermediate frequency (IF) through the mixer (ADMV1018). Instead of an RF phase shifter, the existing on-board local oscillator (LO) phase shifter (ADAR1000) is used for lower cost. The IF signals are delayed by 16 delay lines separately and then combined before being fed to a spectrum analyzer and digital scope. 

\figref{fig:demoJPTAoff} and \figref{fig:demoJPTAon} show the link measurement results. The link distance is 1.2 meters, and the TX power is backed off accordingly to support the demodulated signals' SNR of around $24$ dB. Due to the limited number of delay line ICs we can assemble during the test, only the 8 RX channels on the left are used in this demonstration. This results in lower beamforming gain and correspondingly higher sidelobe than the test using all 16 channels. However, the JPTA functionalities have still been successfully demonstrated. When all the IF delays are set to be equal, the gNB RX works similarly to the traditional analog beamforming. At one time, only one beam can be formed. \figref{fig:demoJPTAoff} shows the case when the RX beam is steered to UE4. The measured beamwidth using 8-channel beamforming is about $10^{\circ}$. Due to the lower beamforming gain at other directions (UE 1-3), only UE4 can be demodulated, and its EVM is $-22.5$ dB.

\figref{fig:demoJPTAon} shows the frequency-dependent multi-beam beamforming results when the IF delays and LO phase shifters are set according to the type-1 codebook. 4 UEs can be supported simultaneously, each occupying $100$ MHz bandwidth. The measured EVM results for UE 1-4 are about $0.5$-$3$ dB worse than those in the traditional analog beamforming method. This matches the $1$-$3$ dB beamforming gain degradation predicted in the Type-1 codebook design. The asymmetrical RX antenna environment and inaccuracy in the analog beamforming alignment also contribute to this difference. Because JPTA relies on full-array beamforming, the $1$-$3$ dB beamforming gain degradation will become less critical as the array size increases.

\section{Simulation Results} \label{sec:Simulation}

\subsection{Simulation Setup}
\begin{figure}[t]
    \centering
    \includegraphics[width= 0.9\linewidth]{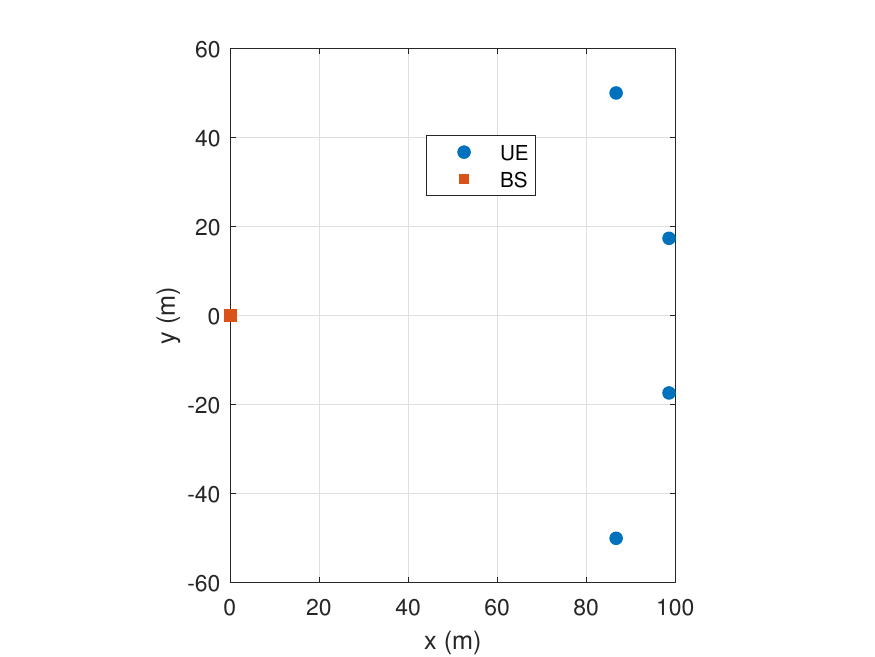}
    \caption{An example network deployment, where four UEs are placed on a ring at a distance 100 m to BS.}
    \label{fig:networkDep}
\end{figure}

\begin{table}[t]
	\centering
	\caption{System level simulation setup}
	\label{tb:setup}
	{%
		\begin{tabular}{|l|l|}
			\hline
			Parameter & Default Value \\ \hline
			Bandwidth & 400 MHz \\ \hline
			Carrier frequency & 28 GHz \\ \hline
			FFT size &  4096 \\ \hline
			Sub-carrier spacing & 120 kHz \\ \hline
			UE transmit power &  23 dBm \\ \hline
			UE beam gain & 0 dB \\ \hline
			BS peak beam gain & 28 dB \\ \hline
			number of analog beams in PAA & 16 \\  \hline
                BS noise figure & 5 dB \\ \hline
                Path-loss exponent & 3 \\ \hline
                TTD delay stepsize & 2.5 ns \\ \hline
	\end{tabular}}
\end{table}

\begin{figure*}[t]
        \centering
        \subfigure[2 UEs.]{
            \includegraphics[width= 0.4\linewidth]{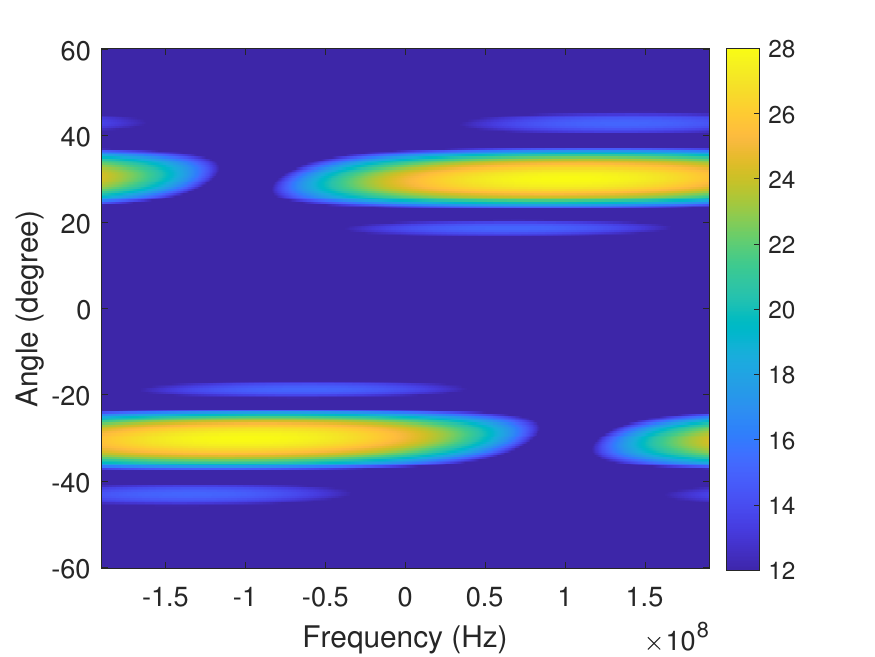}
            \label{fig:JPTA2}}
        \subfigure[4 UEs.]{
            \includegraphics[width= 0.4\linewidth]{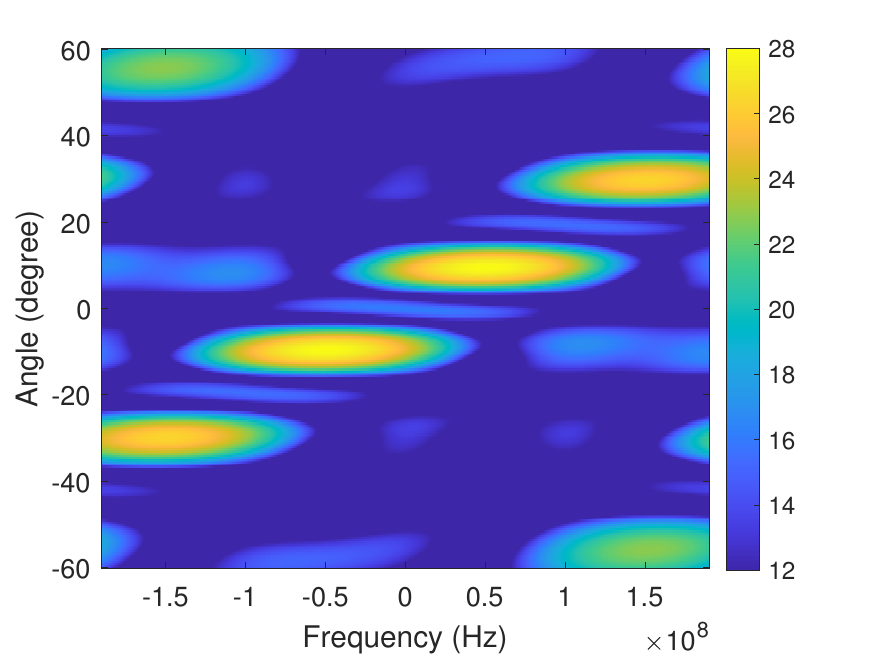}
        \label{fig:JPTA4}}
            \subfigure[8 UEs.]{
            \includegraphics[width= 0.4\linewidth]{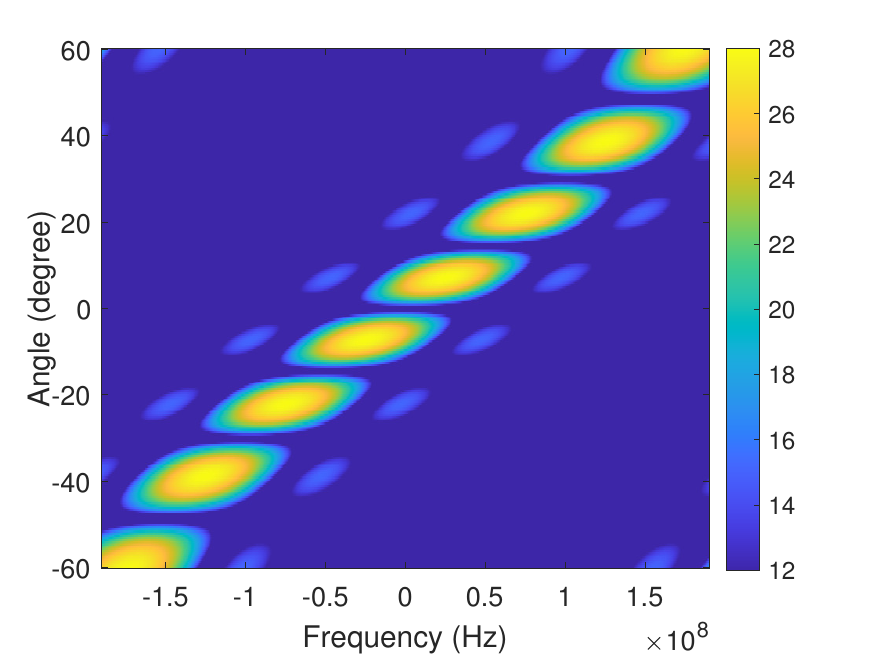}
        \label{fig:JPTA8}}
            \subfigure[16 UEs.]{
            \includegraphics[width= 0.4\linewidth]{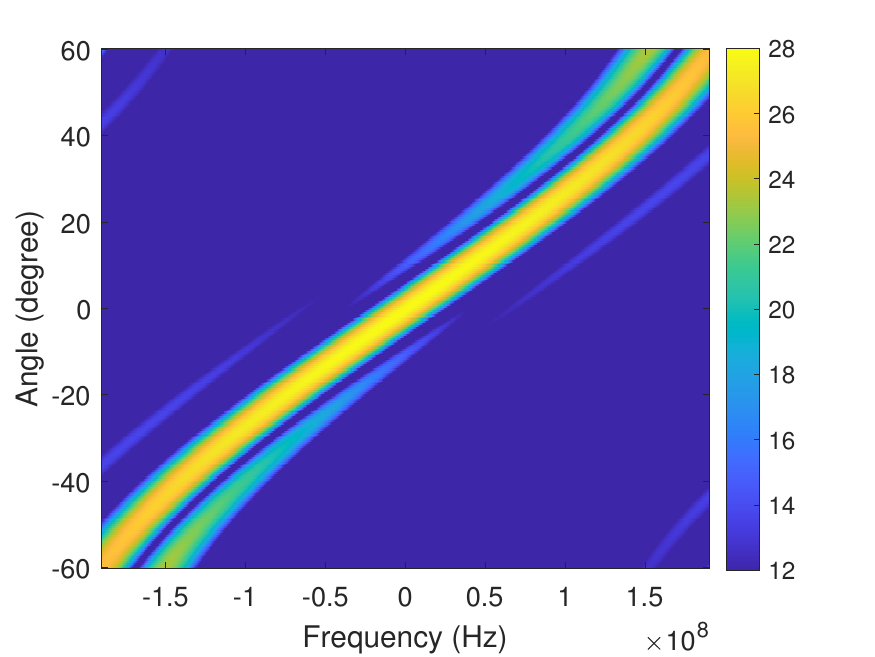}
        \label{fig:JPTA16}}
         \caption{JPTA Type-2 beam pattern for different numbers of UEs. Across the total 400 MHz bandwidth, JPTA generates multiple beams to serve multiple users simultaneously.}
        \label{fig:JPTA_Beams}
\end{figure*}

\begin{figure}[t]
    \centering
    \includegraphics[width= 0.9\linewidth]{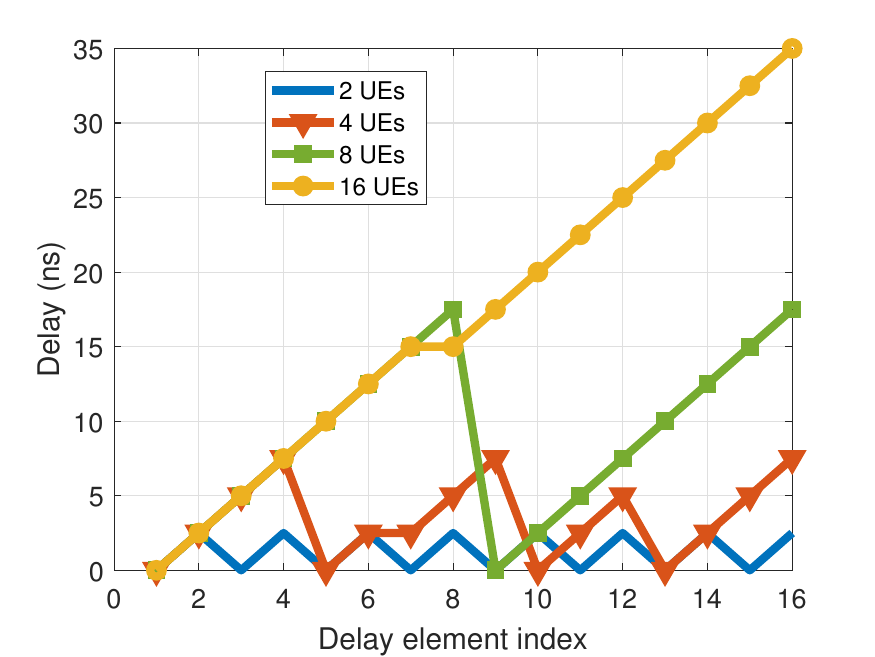}
    \caption{Delay in (ns) for each delay element needed for JPTA for different number of scheduled UEs.}
    \label{fig:JPTADelays}
\end{figure}

\begin{figure}[t]
    \centering
    \includegraphics[width= 0.9\linewidth]{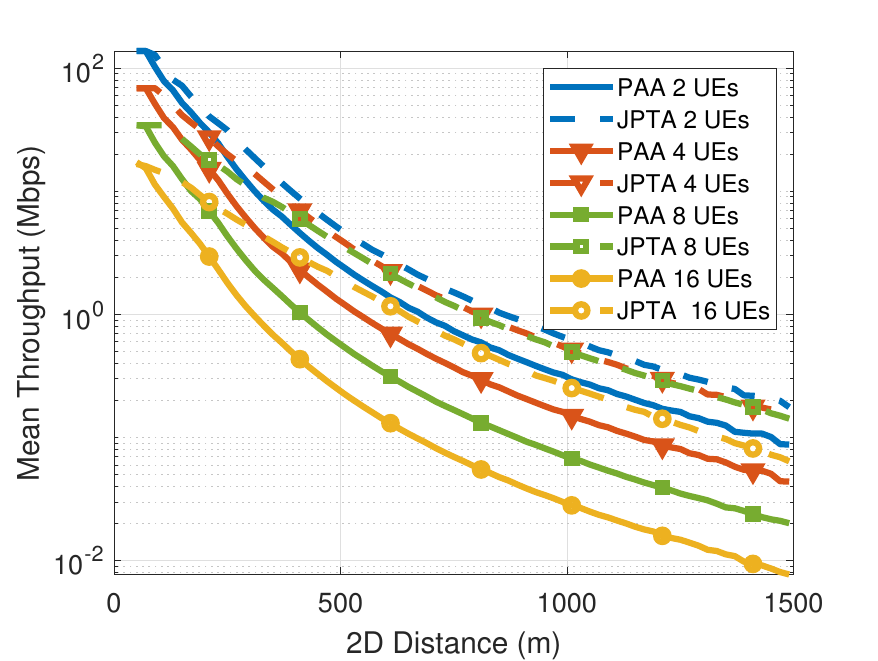}
    \caption{Mean UE throughput vs distance for different number of scheduled UEs. The mean is taken over all the UEs on the same ring.}
    \label{fig:throughput_vs_distances}
\end{figure}

Our simulation setup involves placing the UEs on a ring at fixed distances from the Base Station BS, with the number of UEs varying in each scenario. We consider multiple rings positioned at varying distances from the BS to examine network performance with different serving distances. An illustrative example is presented in  \figref{fig:networkDep}, where four UEs are located per ring at angles $-30^\circ, -10^\circ, 10^\circ,$ and $30^\circ$ relative to the BS at the origin. The rings are positioned at distances ranging from $30$ to $1500$ meters from the BS, but only one ring is active at any given time.

In this paper, we solely consider the effects of large-scale fading. The following equation describes the distance-dependent path gain (in dB):

\begin{align}\label{eq:PLModel}
G_h = 20 \log_{10} \left(\frac{c}{4 \pi f_c} \right) - 10 \beta \log_{10} \left( d \right),
\end{align}
where $c$ represents the speed of light, $f_c$ is the carrier frequency, $d$ denotes the distance between the UE and the BS, and $\beta$ corresponds to the path loss exponent.

During the simulations, we first determine the angle of arrival (AoA) and angle of departure (AoD) in the uplink transmission. Then, we calculate the beam gain for each beam and sub-carrier. In the case of traditional mmWave systems, the beam with the highest gain is selected as the serving beam, and time division multiplexing (TDM) is employed to serve different UEs, with each UE having access to all the resource blocks (RBs). For JPTA, each UE is allocated $\frac{1}{N_{\textrm{UE}}}$ of the total RBs, but all UEs can transmit in every uplink slot. Subsequently, we compute the SNR per sub-carrier per UE, which is converted into the effective SNR using exponential effective SNR mapping (EESM). The effective SNR is mapped to the block error rate (BLER) and throughput for each MCS level, following the approach outlined in \cite{alammouri_extending_2022}. Finally, for each UE, we select the MCS level and number of RBs that maximize the throughput, with the constraint that the BLER remains below $10\%$, and a minimum of four RBs is used.

The BS setup is based on a mmWave antenna unit, featuring a 16V $\times$ 16H array, where each column of antennas is connected to a delay element. The noise figure is 5 dB. The total transmission power is 23 dBm on the UE side, and the beamforming gain is 0 dB. The simulation parameters are summarized in Table \ref{tb:setup}.

Next, we show the system-level simulation results.
The default path loss exponent $\beta$ is set as 3, which is close to the 3GPP urban-micro close-in model whose path loss exponent is ($\beta=3.19$) and the weighted average NLOS path loss exponent ($\beta=2.96$) from the measurement done in 28, 38, 73, 142 GHz \cite{Xing_Yunchou_CL21}. Note that although we do not show the LOS results here, there are more coverage extension benefits brought by JPTA when $\beta=2$ since $K^{1/2}>K^{1/3}$ if $K>1$, according to our analysis in \cite{alammouri_extending_2022}.

\subsection{Simulation Results and Discussion}\label{sec:SimResDis}

Our study focuses on four specific cases: 2 UEs positioned at $[-30^\circ,30^\circ]$, 4 UEs positioned at $[-30^\circ, -10^\circ, 10^\circ,30^\circ]$, 8 UEs uniformly distributed between $[-55^\circ,55^\circ]$, and 16 UEs uniformly distributed between $[-55^\circ,55^\circ]$. The corresponding beam patterns generated by the JPTA are illustrated in \figref{fig:JPTA_Beams}. It is important to note that there is no inter-user interference in these cases since the UEs are assigned non-overlapping sets of RBs. For example, in the case of two UEs, the first half of the RBs are allocated to the UE at $-30^\circ$ and the second half to the UE at $30^\circ$. Additionally, the beam gain varies across different RBs and can decrease up to 3 dB compared to the maximum beam gain of 28 dB, potentially impacting performance. Moreover, it is worth mentioning that the beam pattern for the 16 UEs case closely resembles the Type-2 JPTA beam. Generally, with many UEs dispersed throughout the sector, the Type-1 JPTA beam converges towards a Type-2 beam, often called a rainbow beam.

The corresponding configuration of the delay elements is shown in \figref{fig:JPTADelays}. Note that the larger the required beam coverage, the larger the needed maximum delay. However, as the figure shows, the maximum delay required for the 2 UEs, 4 UEs, 8 UEs, and 16 UEs cases are $2.5$, $7.5$, $17.5$, and $35$ ns, respectively. 

The performance improvement achieved by JPTA can be observed in \figref{fig:throughput_vs_distances}, which depicts the average throughput per UE at various serving distances for the four scenarios discussed. The solid lines represent the benchmark scenario of a traditional mmWave beam sweeping by a phased antenna array.

When UEs are located very close to the BS, they experience sufficiently high SNRs that enable them to utilize the highest MCS level across the entire bandwidth. Consequently, in the case of PAA, UEs at these short distances employ the highest MCS with the maximum number of RBs. However, due to the antenna sweeping procedure, each UE can only transmit every $N_{\textrm{UE}}$ time slots. Similarly, UEs with high SNRs in the JPTA scheme can also use the maximum MCS level. Although they have a smaller portion of RBs available, specifically $\frac{1}{N_{\textrm{UE}}}$ of the total RBs, they are able to transmit on every UL time slot. Consequently, the performance of JPTA and PAA is identical at very short serving distances, where using $N$ times the BW is equivalent to having $N$ UL time slots for extremely high SNRs. Conversely, when the serving distance becomes sufficiently large, e.g., at cell-edge, UEs in both schemes operate at MCS level 0 and utilize 4 RBs. However, each UE in JPTA can transmit on every UL time slot, resulting in JPTA achieving a throughput that is $N_{\textrm{UE}}$ times higher than that of PAA.

In summary, JPTA does not yield any throughput enhancement at short distances, while the gain increases to $N_{\textrm{UE}} \times 100\%$ at the cell edge. In regions with moderate SNRs, JPTA consistently outperforms PAA significantly. For instance, in the case of 16 UEs and at a distance of 1500 m, JPTA delivers an impressive throughput enhancement of $830\%$ compared to the traditional PAA. This improvement can be attributed to the fact that having $N_1$ UL time slots with $\frac{N_2}{k}$ RBs (as in JPTA) is always superior to having $\frac{N_1}{k}$ UL time slots with $N_2$ RBs (as in PAA), for any integer $k$, as the latter case leads to $k$ times higher noise power.

An alternative perspective to assess the benefits of JPTA can be gained by considering throughput coverage, which refers to the maximum distance at which the throughput exceeds a specific threshold. 
For instance, when considering a target throughput of $1$ Mbps for a scenario with 8 UEs, only UEs located within a distance of $410$ m can achieve the desired throughput when using PAA. However, when JPTA is employed, this coverage distance increases to $790$ m. This ratio is close to $N_{\textrm{UE}}^{\frac{1}{3}} = 8^{\frac{1}{3}}=2$ as elaborated in \cite{alammouri_extending_2022}.

\section{Conclusion}\label{sec:Conclusion}

In this paper, we introduced a novel technology called Joint Phase-Time Array (JPTA), which enables the generation of multiple beams in the frequency domain without requiring multiple RF chains. Our study encompassed various aspects, including beam generation, hardware feasibility, and performance evaluation. The simulation results demonstrated the significant benefits in uplink transmission offered by JPTA, including extending the uplink throughput coverage. Furthermore, JPTA only needs much less cost than the current mmWave MIMO technology. Therefore, We believe that JPTA is a promising candidate for next-generation mmWave MIMO technology.

\bibliographystyle{IEEEtran}
\bibliography{AlAmmouri.bib,Vishnu.bib,TTDRef.bib}
\end{document}